\documentclass[conference,usenatbib]{basi}
\usepackage[T1]{fontenc}
\usepackage[british]{babel}
\usepackage[varg]{txfonts}
\usepackage{float}
\usepackage{rotating}
\usepackage{dcolumn}
\begin{document}
\title[Magnetohydrodynamic stability of stochastically driven accretion flows]{Magnetohydrodynamic stability of stochastically driven accretion flows}
\author[Sujit~Kumar~Nath]%
       {Sujit~Kumar~Nath$^1$\thanks{email: \texttt{sujitkumar@physics.iisc.ernet.in}},
      \\
       $^1$Department of Physics, Indian Institute of Science, Bangalore 560012, India}
\pubyear{2013}
\volume{**}
\pagerange{**--**}

\date{Received --- ; accepted ---}

\maketitle
\label{firstpage}

\begin{abstract}
We investigate the evolution of magnetohydrodynamic perturbations in presence of stochastic noise
in rotating shear flows. The particular emphasis
is the flows whose angular velocity decreases but specific angular momentum increases with increasing
radial coordinate. Such flows, however, are Rayleigh stable, but must be turbulent in 
order to explain astrophysical observed data and, hence, reveal a mismatch between the linear theory and 
observations/experiments. The mismatch seems to have been resolved, atleast in certain regimes,
in presence of weak magnetic field revealing magnetorotational instability. 
The present work explores the effects of stochastic noise on such magnetohydrodynamic flows, in order 
to resolve the above mismatch generically for the hot flows. It is found that such stochastically driven flows exhibit large temporal and spatial auto-correlations and cross-correlations of perturbation and hence large energy dissipations of perturbation, which generate instability. 
\end{abstract}


\section{Introduction}\label{s:intro}
In the present study, we implement the ideas of statistical physics, already implemented by 
Mukhopadhyay \& Chattopadhyay (2013), to rotating, magnetized, shear flows 
in order to obtain the correlation
energy growths of perturbation and underlying scaling properties. We essentially concentrate on a small section of such a flow which is nothing but a plane shear 
flow supplemented by the Coriolis effect, mimicking a small section of an astrophysical accretion disk.
\section{Equations describing perturbed magnetized rotating shear flows in presence of noise}\label{s:fonts}
The linearized Navier-Stokes equation in presence of background plane shear $(0,-x,0)$ and magnetic field $(0,B_1,1)$,
when $B_1$ being a constant and velocity and magnetic field perturbations $(u,v,w)$ and $(B_x,B_y,B_z)$ respectively, in presence of angular velocity 
$\Omega\sim r^{-q}$, in a small section of the incompressible flow, has already been established (\cite{mc13}). 
The underlying equations
are nothing but the linearized set of hydromagnetic equations including the equations of induction 
in a local Cartesian coordinate. These equations supplemented by conditions of incompressibility and absence of magnetic charge can be recasted into magnetized version of Orr-Sommerfeld and Squire equations in presence of the Coriolis force and stochastic noise, given by
\begin{equation}
\left(\frac{\partial}{\partial t}-x\frac{\partial}{\partial y}\right)\nabla^2 u
+\frac{2}{q}\frac{\partial \zeta}{\partial z}-\frac{1}{4\pi}\left(B_1\frac{\partial}{\partial y}+\frac{\partial}{\partial z}\right)\nabla^2B_x
=\frac{1}{R_e}\nabla^4 u+\eta_1(x,t),
\label{orrv}
\end{equation}
\begin{equation}
\left(\frac{\partial}{\partial t}-x\frac{\partial}{\partial y}\right)\zeta
+\frac{\partial u}{\partial z}
-\frac{2}{q}\frac{\partial u}{\partial z}-\frac{1}{4\pi}\left(B_1\frac{\partial}{\partial y}+\frac{\partial}{\partial z}\right)\zeta_B=\frac{1}{R_e}\nabla^2 \zeta +
\eta_2(x,t),
\label{zeta}
\end{equation}
\begin{eqnarray}
\left(\frac{\partial }{\partial t}-x\frac{\partial}{\partial y}\right)B_x
-B_1\frac{\partial u}{\partial y}-\frac{\partial u}{\partial z}=
\frac{1}{R_m}\nabla^2B_x+\eta_3(x,t),
\label{orrb}
\end{eqnarray}
\begin{eqnarray}
\left(\frac{\partial }{\partial t}-x\frac{\partial}{\partial y}\right)\zeta_B-
\frac{\partial \zeta}{\partial z}-B_1\frac{\partial \zeta}{\partial y}-
\frac{\partial B_x}{\partial z}=\frac{1}{R_m}\nabla^2\zeta_B+\eta_4(x,t).
\label{orrbzeta}
\end{eqnarray}
where $\eta_{1,2,3,4}$ are the components of noise arising in the linearized system due
to stochastic perturbation such that $<\eta_i(\vec x,t) \eta_j(\vec x',t')>=D_i(\vec x)\:\delta^3(\vec x-\vec x')\:\delta(t-t')\:\delta_{ij}$.
The long time, large distance behaviors of the correlations of noise are encapsulated in 
$D_i(\vec x)$ which is a structure pioneered by Forster, Nelson \& Stephen (1977). 
In the Fourier space, however,
the structure of the correlation function $D_i(\vec k)$
depends on the regime under consideration.
It can be shown for all (non-linear) non-inertial flows (\cite{nelson,akc_shear}) that $D_i(k) \sim 1/k^d$, where $d$ is the spatial dimension, without
vertex correction and $D_i(k) \sim 1/k^{d-\alpha}$, with $\alpha>0$, in presence of vertex correction.
Note, however, that $D_i(\vec x)$ is constant for white noise.
Since we are focusing onto the narrow gap limit, we can resort to a Fourier series expansion of
$A$(=$u$, $\zeta$, $B_x$, $\zeta_B$) as
\begin{eqnarray}
\nonumber
A(\vec{x},t)=\int\tilde{A}_{\vec{k},\omega}\,e^{i(\vec{k}.\vec{x}-\omega t)}d^3k\,d\omega,\\
\nonumber
\eta(\vec{x},t)=\int\tilde{\eta}_{\vec{k},\omega}\,e^{i(\vec{k}.\vec{x}-\omega t)}d^3k\,d\omega,\\
\nonumber
\label{four}
\end{eqnarray}
and substituting them into equations (\ref{orrv}), (\ref{zeta}), (\ref{orrb}) and (\ref{orrbzeta}) we obtain a set of four linear equations of the form, given by
\begin{eqnarray}
(\tilde{A}_{\vec{k},\omega})_i=\sum\limits_{j}({\cal N})_{ij}(\tilde{\eta}_{\vec{k},\omega})_j, 
\label{mat1}
\end{eqnarray}
\section{Two-point correlations of perturbation in presence of white noise}
We now look at the spatio-temporal autocorrelations 
of the perturbation flow fields $u$, $\zeta$, $B_x$ and $\zeta_B$ for a very large fluid and 
magnetic Reynolds numbers (\cite{barabasi_stanley}). This choice is quite meaningful for 
astrophysical flows. 
For the present purpose, the magnitudes and gradients (scalings) of these correlations of perturbations would 
plausibly indicate noise induced instability which could lead to 
turbulence in rotating shear flows.
\subsection{Temporal and Spatial correlations}
Assuming $<\tilde{\eta_i}_{\vec{k},\omega}\,\tilde{\eta_j}_{-\vec{k},-\omega}>=\delta_{ij}$, without loss of any important physics,
we obtain the temporal and spatial correlations of perturbations given below as
\begin{eqnarray}
\nonumber
&&<A_i(\vec{x},t)\,A_j(\vec{x},t+\tau)>=C_{A_iA_j}(\tau)=\int d^3k\,d\omega\,e^{-i\omega\tau}<\tilde{A_i}_{\vec{k},\omega}\,
\tilde{A_j}_{-\vec{k},-\omega}>\\
\nonumber
&&<A_i(\vec{x},t)\,A_j(\vec{x}+\vec{r},t)>=S_{A_iA_j}(r)=\int d^3k\,d\omega\,e^{i\vec k.\vec r}
<\tilde{A_i}_{\vec{k},\omega}\,\tilde{A_j}_{-\vec{k},-\omega}>,
\nonumber
\end{eqnarray}
where $A_i=u,\zeta,b_x~or~\zeta_b$. For $i=j$ we get autocorrelations and for $i\neq j$ we get cross-correlations.  We further consider the projected hyper-surface for which $k_x=k_y=k_z=k/\sqrt{3}$, without much loss of
generality for the present purpose.
\begin{figure}[H]
\centerline{\includegraphics[width=6.2cm]{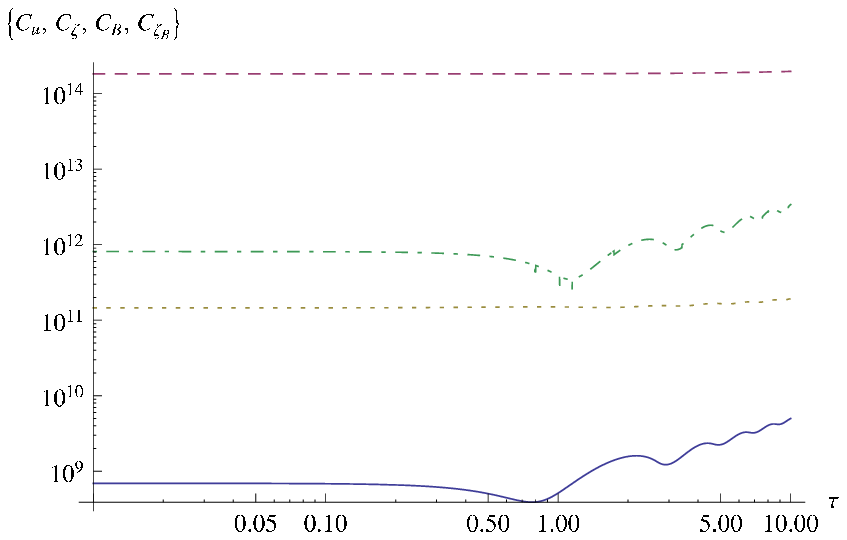} \qquad
            \includegraphics[width=6.2cm]{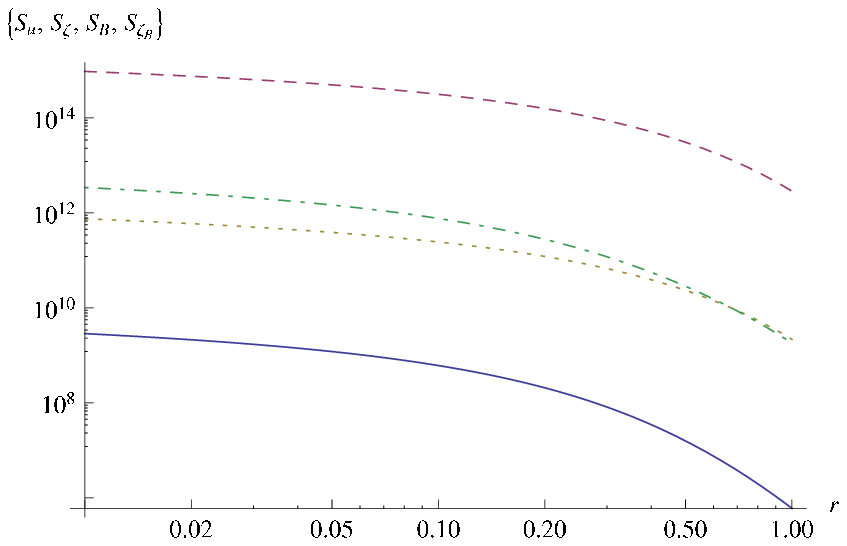}}
\caption{Temporal (left panel) and spatial (right panel) autocorrelations of velocity (solid line), vorticity (dashed line), magnetic field (dotted line) and 
magnetic vorticity (dot-dashed line), when $q=1.5$.
\label{f:auto}}
\end{figure}
From Figure \ref{f:auto} it is evident that flows of above mentioned kind exhibit large temporal and spatial autocorrelations of perturbation and hence large energy dissipations of perturbation at least in the time and length scales of interest, leading to instability and plausible turbulence.
\section{Summary and conclusions}\label{s:ADS}
In this work, we have attempted to address the origin of instability and then turbulence in 
magnetized, rotating, shear flows in presence of stochastic noise. Our particular emphasis is
the flows having decreasing angular velocity but increasing specific angular momentum with the radial
coordinate, which are Rayleigh stable. The flows with such a kind of velocity profile are 
often seen in astrophysics. 
As the molecular viscosity in astrophysical accretion disks
is negligible, any transport of matter therein would arise through turbulence only, in order
to explain observed data. In the cases of hot flows, e.g. disks around black holes, magnetorotational instability is generally 
believed to be responsible for turbulence and hence transport of angular momentum therein. However many authors argued for limitations of magnetorotational instability (\cite{mk,paoletti,avila}).
Therefore, essentially we have addressed here the
plausible origin of viscosity in rotating shear flows of the kind mentioned above.
\section*{Acknowledgements}
I would like to thank Banibrata Mukhopadhyay for suggesting the problem and discussing 
throughout the course of this work. This work was partly supported by the ISRO grant ISRO/RES/2/367/10-11.

\end{document}